\documentclass{aa}
\usepackage{amssymb}
\usepackage{amsmath}
\usepackage{graphicx} 
\usepackage{txfonts}

\newcommand{\reac}[6]{$\mathrm{^{#1}#2(#3,#4)^{#5}#6}$}
\newcommand{\chem}[2]{$\mathrm{^{#1}#2}$}

\begin{document}


\title{The s-process weak component: uncertainties due to convective overshooting}
\titlerunning{The s-process weak component}
\author{V. Costa \inst{1}, M. L. Pumo \inst{2}, A. Bonanno \inst{1},
	and R. A. Zappal\`a \inst{2}
}
\offprints{V. Costa, \email{vcosta@oact.inaf.it}}
\institute{Istituto Nazionale di Astrofisica - Osservatorio Astrofisico di Catania (INAF-OAC),
           Citt\`a Universitaria, Via S. Sofia 78, I-95123 Catania, Italy\\
           \and Universit\`a di Catania, Dipartimento di Fisica e Astronomia (sez. astrofisica),
	   Citt\`a Universitaria, Via S. Sofia 78, I-95123 Catania, Italy
}
\date{Received 1 August, 2005; accepted 29 September, 2005}

\abstract{
Using a new s-nucleosynthesis code, coupled with the stellar evolution code Star2003, we  performed simulations to study the impact of the convection treatment on the s-process during core He-burning in a $25\, M_{\odot}$ star (ZAMS mass) with an initial metallicity of $Z$=$0.02$.
Particular attention was devoted to the impact of the extent of overshooting on the s-process efficiency.
The results show enhancements of about a factor $2\mbox{-}3$ in s-process efficiency (measured as the average overproduction factor of the $6$ s-only nuclear species with $60\lesssim A\lesssim 90$) with overshooting parameter values in the range $0.01\mbox{-}0.035$, compared to results obtained with the same model but without overshooting. The impact of these results on the p-process model based on type II supernovae is discussed.

}

\maketitle
\section{Introduction}
\label{introduction}

As already pointed out by \cite{clayton}, it is now widely accepted that s-nuclei are formed via neutron exposures on iron-peak nuclei, and it is also clear that more than one event (or s-process ``component'') with a single set of physical conditions (like neutron exposure $\tau$, initial abundances, and neutron density) is necessary in order to account for the observed solar distribution of s-nuclei abundances.\par
Current views on the subject suggest the existence of two components. One is the so-called ``main'' component that is associated with low-intermediate mass stars ($\sim 1\mbox{-}5\, M_{\odot}$) during their AGB phase and that should be responsible for the synthesis of s-species from $A=90$ up to \chem{209}{Bi} (Gallino et al. \cite{gallino98}; Busso et al. \cite{busso99}, \cite{busso01}; Herwig et al. \cite{herwig03}; Lugaro et al. \cite{lugaro03}; Goriely and Siess \cite{gorielysiess2004}).
The other is the so-called ``weak'' component that covers the $60 \lesssim A \lesssim 90$ mass range, and is found in massive ($M \gtrsim 10\, M_{\odot}$) stars during their core He-burning phase, as first proposed by Peters (\cite{peters}).
Moreover, a particular class of massive stars that may contribute to the s-process enrichment of interstellar medium includes the so-called {\it Wolf-Rayet} stars (Meynet et al. \cite{meynet2001} and references therein). Thanks to the loss of their initial H envelopes, these stars may show the products of He-burning on their surface.\par
Some authors (Gallino et al. \cite{gallino98}; Busso et al. \cite{busso99}, K\"appeler \cite{kappeler99}; Goriely and Siess \cite{gorielysiess2001}, \cite{gorielysiess2004}; Lugaro et al. \cite{lugaro03}) also suggest the existence of a ``strong'' component that should be responsible for the synthesis of s-species around \chem{208}{Pb}. This component should occur in low-metallicity stars of low-intermediate mass during the AGB phase.  Observational evidence of an enhanced lead abundance in metal-poor stars (Van Eck et al. \cite{vaneck2001}) seems to confirm this possibility.\par
Although the general features of s-nucleosynthesis seem to be well-established, there are still ambiguities and uncertainties concerning the weak component. Some authors (see e.g. Raiteri et al. \cite{raiteri93};  The et al. \cite{theetal1}; Hoffman et al. \cite{hoffman01}, Woosley et al. \cite{woosley02}) suggest that this process might be prolonged during subsequent post-He-burning stellar evolutionary phases, so that the s-abundances at the end of core He-burning could be modified, but this hypothesis needs further investigation (Arcoragi et al. \cite{arcoragi91}; The et al. \cite{theetal1}; Rayet \& Hashimoto \cite{rayet2}, Woosley et al. \cite{woosley02}).
Besides, both the nuclear mass range listed above for the weak component and the overall efficiency of the process are subject to significant uncertainties that come from both the nuclear physics and astrophysical modelling aspects of the problem.\par
Uncertainties due to nuclear physics are connected to the value of the  \reac{12}{C}{\alpha}{\gamma}{16}{O} reaction rate, with its impact on the stellar core structure, and to reaction rates on which the so-called ``neutron economy'' (the balance between neutron emission and captures) is based. Many authors (see e.g. Rayet \& Hashimoto \cite{rayet2}; The et al. \cite{theetal1}; Rayet et al. \cite{rayet3}; Costa et al. \cite{costa1}, \cite{costaiapichino}, hereafter Paper I and Paper II) have examined the impact of these uncertainties on the s-process weak-component yields.
On the astrophysical side, the lack of a self-consistent convection theory is still one of the major problems in  calculating stellar structures. Thus it can be a significant source of uncertainty for nucleosynthesis processes, like the s-process weak component.\par
Some weak component s-process calculations performed so far have been developed using stellar evolution codes that are based on simple mixing length theory without overshooting (see e.g. Rayet \& Hashimoto \cite{rayet2}; The et al. \cite{theetal1}). However, overshooting may affect s-process nucleosynthesis in massive stars by giving rise to an increase in the convective core mass.
This may affect the s-process through lengthening of the He-burning lifetime and an increase in the amount of material that experiences neutron irradiation.
Indeed, matter originally lying in stellar regions with temperatures that are too low to allow the main neutron source to work (\reac{22}{Ne}{\alpha}{n}{25}{Mg} becomes effective at temperatures $T\gtrsim 2.5\times 10^8 ~K$) would be mixed towards central hotter regions, while a change of the He-burning lifetime would directly change the s-process lifetime.\par
Langer et al. (\cite{langer89}) used two $30\, M_{\odot}$ star models (without and with overshooting) to study the overshooting effect on the s-process in massive stars.
They conclude that overshooting leads to an increase in s-nuclei production (see their Table 2) with an enhancement factor of about $1.5\mbox{-}2$ on final mass fractions and an increase in the average number of captured neutrons per nucleus of \chem{56}{Fe} from a value of $3.6$ to a value of $4.5$.\par
We re-examined the impact of overshooting on the s-process not only because a comprehensive study of the role of overshooting on s-nucleosynthesis is needed in the framework of recent nuclear data and stellar models, but also because the approach of Langer et al. (\cite{langer89}) for overshooting is based on a simple artificial enhancement of the convective core extension by a $40\%$ fraction of pressure scale height, while we used a more sophisticated approach based on a diffusive algorithm.
In this approach an ``overshooting parameter'' $f$ determines the extension of convectively mixed core, with the overall efficiency of overshooting described by a diffusion equation (see Sect. \ref{stellarmodel} for more details). This approach has  already been tested with some success for the Sun in order to find a solution to the lithium depletion puzzle (see e.g. Bl\"ocker et al. \cite{blocker}; Schlattl \& Weiss \cite{schlattl99}). Of course the overshooting in massive stars is different, as it is concerned with the ``outer'' border of the central convection zone, while the Sun is characterised by overshooting through the ``inner'' border of the external convection zone.
However, Salasnich et al. (\cite{salasnich_chiosi99}) investigated the internal convection zone in massive stars using diffusion to model the convective overshooting. They give some indications of how to modify the overshooting parameter in order to account for the observed distribution of massive stars across the HR diagram, and, in particular, they claim a value of about $f=0.015$ for $20\, M_{\odot}$ stellar models for a setting of this parameter. Other authors (Young et al. \cite{young01}) suggest that the overshooting parameter may be mass-dependent and that massive stars probably require a higher $f$ value. Thus, since the proper $f$ value is still being debated, we examine a whole range of its possible values in this paper.\par
Moreover, our study may be of interest in p-process modellings, because some of the problems in the model for the p-process (taking place in the type II supernovae O-Ne layers) might be linked to uncertainties in the s-process weak component efficiency (Paper I; Paper II; Rayet et al. \cite{rayet3}; Arnould \& Goriely \cite{Arnould03}), since the relevant nuclides are p-process seeds.

\section{Input physics}
\subsection{Stellar model}
\label{stellarmodel}

A preliminary set of nucleosynthesis calculations was performed with the same stellar data as in Papers I and II. It is basically the model of a He star of $8\, M_{\odot}$ taken from Nomoto \& Hashimoto (\cite{nomotohashimoto}), which is supposed to correspond to a $25\, M_{\odot}$ at the ZAMS. This model treats convection with the standard mixing length theory and does not include overshooting or mass loss.\par
Our new calculations were performed with the stellar evolution code Star2003, with the Garching stellar evolution code (Weiss \& Schlattl \cite{Weiss00}) as its basis, and the models were evolved starting from ZAMS until the end of core He-burning.\par
Mixing was modelled within a diffusion approach (Bl\"ocker et al. \cite{blocker}, Freytag et al. \cite{freytag}), and abundance changes of the 12 considered nuclear species were calculated with the following diffusion equation:
\begin{equation}
\label{diffusion}
\frac{dX}{dt} = \left(\frac{\partial X}{\partial t}\right)_{nuc} + \frac{\partial}{\partial m_r}
\left[\left(4\pi r^2 \rho\right)^2 D \frac{\partial X}{\partial m_r} \right]_{mix},
\end{equation}
where the first term on the right is the time derivative of a given nuclidic mass fraction due to nuclear reactions, while the second is the diffusive term that describes mixing. The derivatives were calculated with respect to the radial stellar mass $m_r$, while the other symbols used in Eq. \ref{diffusion} have their standard meaning. The difference among convective, overshooting, and radiative regions lies in the value used for the diffusion coefficient $D$. In convective zones, the diffusion coefficient is given by:
\begin{equation}
\label{diffusion_coefficient_1}
D_{conv} = \frac{1}{3} v_c l,
\end{equation}
where $v_c$ is the average velocity of convective elements derived from the mixing length theory, and $l$ is the mixing length.
For the overshooting regions, the following diffusion coefficient was used instead:
\begin{equation}
\label{diffusion_coefficient_2}
D_{over} = D_0 exp \frac{-2z}{H_v}; \; ~~~~~~~~~~ H_v = f\cdot H_p
\end{equation}
where $D_0$ is taken as equal to the value of $D_{conv}$ at the upper radial edge of the convection zone established through the Schwarzchild criterion; $z = |r - r_{edge}|$ is the radial distance from the same edge; $H_p$ is the pressure scale height, while $f$ is the overshooting parameter that determines the overall efficiency of overshooting and the extension of the convectively mixed core.
For radiative zones, the diffusion coefficient was put to zero, and abundance changes were only due to the nuclear reaction term.\par
As for energy generation and chemical evolution, the code includes a hydrogen burning network (pp chain, CNO cycle) and a He-C-O network, following the evolution of 12 nuclides (\chem{1}{H}, \chem{3}{He}, \chem{4}{He}, \chem{12}{C}, \chem{13}{C}, \chem{14}{N}, \chem{15}{N}, \chem{16}{O}, \chem{17}{O}, \chem{20}{Ne}, \chem{24}{Mg}, \chem{28}{Si}).
Reaction rates were taken from Caughlan et al. (\cite{cf85}). The rate of \reac{12}{C}{\alpha}{\gamma}{16}{O} is about $20\%$ higher than the one reported in the NACRE  compilation (Angulo et al. \cite{nacre}), which is also the one included in the s-process reaction network described in Sect. \ref{the_reaction_network}.
This choice was made in order to make our results comparable to those obtained in Papers I and II. Note that the \reac{12}{C}{\alpha}{\gamma}{16}{O} rate from Caughlan et al. (\cite{cf85}) is inside the uncertainty range reported in NACRE.
Calculations entirely based on updated rates will be presented in a future paper.\par
As for the equation of state, the OPAL01 EOS tables are used, which are an updated version of the OPAL96 tables, revised according to the prescriptions by Rogers (\cite{rogers01}) (see Bonanno et al. \cite{bonanno02} for details). The OPAL opacity (Iglesias \& Rogers \cite{iglesias96}) are used, implemented in the low-temperature regime by the tables of Alexander \& Ferguson (\cite{alex94}).

\subsection{The nucleosynthesis code}
\label{thenucleosynthesiscode}
Our new s-nucleosynthesis code uses a differential equation solver that is based on the implicit Kaps-Rentrop method, which provides an estimate of the ``truncation error'' associated with each single integration step (see Press et al. \cite{numrecipies} for details). In this way settings concerning  precision requests can be based on truncation error constraints. Some numerical tests on the Kaps-Rentrop method and a comparison with other numerical solving procedures can be found in Timmes (\cite{timmes}). The s-process calculations reported by Papers I and II, Rayet et al. (\cite{rayet1}), and Rayet \& Hashimoto (\cite{rayet2}) were based on the Wagoner two-step linearisation procedure (Wagoner \cite{wagoner69}).
The Wagoner method does not provide any error estimate, so that precision is only based on putting a limit on the relative variations of isotopic abundances from one time step to the next.\par
The ``post-processing'' technique is used to couple the nucleosynthesis simulations with stellar evolution data.
This technique can easily be applied to radiative zones, where nucleosynthesis occurs according to local physical conditions. 
The prescriptions described by Prantzos et al. (\cite{prantzos87}) are used for the convective core, 
but as they do not say how to include the effects from changes in the convection zone borders' position with time, we added the prescription described by Kippenhahn \& Weigert (\cite{kippenan}) to include this effect.\par
If $\dot{X}$ is the ``local'' rate of change of the mass fraction of a given nuclide (due to local physical conditions), $\dot{m_1}$ and $\dot{m_2}$ are the time derivatives of convection zone borders' position (mass coordinates), $M_c = (m_2-m_1)$ is the convection zone mass extension, and $X_1$ and $X_2$ the mass fractions of the same nuclide just below and above the convection zone, as a mass fraction discontinuity can generally be present in convection zone borders. Then if $\chi$ is the average mass fraction uniformly valid  over the whole convection zone for the given nuclide, then the rate of change of $\chi$ may be expressed as follows:
\begin{equation}
\label{aver_mass_frac_change}
\dot{\chi} = \frac{1}{M_c}\left[ \int_{m_1}^{m_2} \dot{X}dm + (\chi-X_1)\dot{m_1} +
(X_2-\chi)\dot{m_2} \right].
\end{equation}
The first term on the right needs the evaluation of ``averaged'' rates considered by Prantzos et al. (\cite{prantzos87}), while the second and third terms were implemented in our code in order to treat  changes in the convection zone borders.
The second term on the right side of Eq. \ref{aver_mass_frac_change} is considered only when $\dot{m_1} < 0$, while the third term is considered only when $\dot{m_2} > 0$, which means that these two terms take into account the inclusion of material that is not previously mixed into the convection zone when it expands.
That is why the two terms above are not considered when the convection zone shrinks.\par
Another aspect that has to be taken into account when coupling the stellar model with the s-nucleosynthesis code is related to the mixing inside the convective core. While the stellar evolution code uses diffusion to model the mixing of the species involved in the 12-nuclide network described above (Sect. \ref{stellarmodel}), instantaneous mixing is assumed for all s-nuclides where a local equilibrium is assumed to hold for the neutrons (see Prantzos et al. \cite{prantzos87} for details).
Thus the overall procedure is not strictly self-consistent; but we believe that, while the inclusion of diffusion in the stellar code allows us to model the stellar core structure evolution better (in terms of temperature profiles, mass extension, He-burning lifetime), neglecting diffusion in the nucleosynthesis code should not have important consequences, as long as the wide uncertainties in stellar structure due to convection uncertainties are under study.
A more refined and self-consistent approach will be presented in a future paper.

\subsubsection{The reaction network}
\label{the_reaction_network}
The s-nucleosynthesis code uses a much wider nuclear network than the one used by the stellar evolution code.
This network is the same as the one in Papers I and II, Rayet \& Hashimoto (\cite{rayet2}), and Rayet et al. (\cite{rayet3}). It includes $472$ nuclides and $834$ reactions. Differences are present in the reaction rates; in particular, rates from the {\it Network Generator} NETGEN\footnote{Available online at http://www.astro.ulb.ac.be/Netgen/} (Aikawa et al. \cite{Netgen}) were used.\par
In order to test the dependability of the s-nucleosynthesis code, a preliminary s-process calculation was  performed with the same rates and stellar model as in Papers I and II, since our results (see Table \ref{tab:old_model}) reproduced theirs quite well; we refer to the results obtained by them with NACRE {\it adopted} values for \reac{22}{Ne}{\alpha}{n}{25}{Mg} and \reac{22}{Ne}{\alpha}{\gamma}{26}{Mg} reaction rates.

\section{Models}
\label{Models}

To estimate the impact of convective overshooting on the s-process efficiency, we calculated the evolution of a $Z=0.02,~25\,M_{\odot}$ star\footnote{Initial composition was taken from Anders \& Grevesse (\cite{anders_grevesse89}).} with the code Star2003 and with mixing length parameters $\alpha=1.7$ and $2.0$ and, for overshooting parameters, $f=10^{-5}$ (model without overshooting), $0.01, 0.015, 0.02,$ $0.025, 0.03$, and $0.035$.\par
\begin{figure}
  \centering
  \includegraphics[height=8cm]{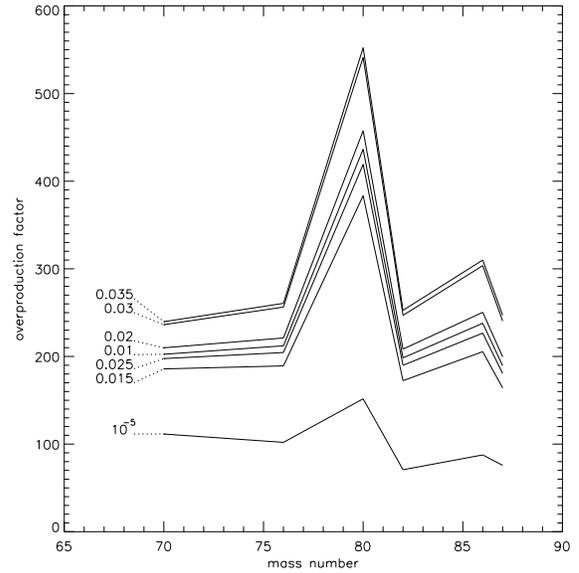}
  \caption{Overproduction factors of the $6$ s-only nuclei within the mass range $70\leq A\leq 90$ as a function of mass number for stellar models with mixing length parameter $\alpha=1.7$. The labels refer to the overshooting parameter value.}
  \label{fig:Fi-a1.7}
\end{figure}
\begin{figure}
  \centering
  \includegraphics[height=8cm]{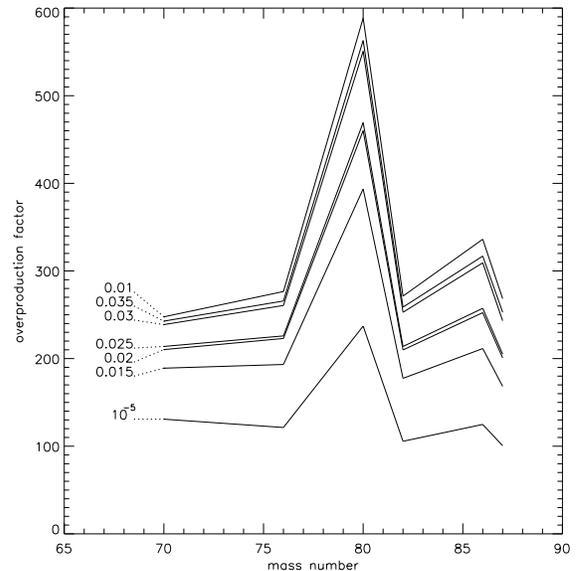}
  \caption{Same as Fig. \ref{fig:Fi-a1.7} but for stellar models with mixing length parameter $\alpha=2.0$.}
  \label{fig:Fi-a2.0}
\end{figure}

The s-process efficiency can be analysed in terms of the following parameters:
\begin{itemize}
 \item[-] the average overproduction factor $F_0$ for the $6$ s-only nuclei \chem{70}{Ge}, \chem{76}{Se}, \chem{80}{Kr}, \chem{82}{Kr}, \chem{86}{Sr}, and \chem{87}{Sr} given by $$F_0=\frac{1}{N_s}\sum_{i}F_i \mbox{\,~~~~~~ with\,~ } F_i=\frac{X_i}{X_{i,ini}},$$ where $F_i$ is the overproduction factor, $X_i$ the mass fraction (averaged over the convective He-burning core) of s-only nucleus i at the end of s-process, $X_{i,ini}$ the initial mass fraction of the same nucleus, and $N_s=6$ is the number of s-only species considered;
 \item[-] the average overproduction factor $\tilde{F_0}$ identical to $F_0$, but for the $18$ s-nuclei
 \footnote{
 These are the nuclei indicated as ``mainly'' produced by the s-process in Anders \& Grevesse (\cite{anders_grevesse89}).
 }
 within the mass range $88\leq A \leq 130$;
 \item[-] the maximum mass number $A_{max}$, for which the species in the $60 \leq A \leq A_{max}$ mass range are overproduced by at least a factor of about $10$ and $5$ (respectively, the first and second value in the fourth column of Tables \ref{tab:alpha1.7} and \ref{tab:alpha2.0}, and first and second value in the third column of Table \ref{tab:old_model});
 \item[-] the number of neutrons captured per \chem{56}{Fe} seed nucleus $n_c$, defined as $$n_c=\sum_{A=57}^{209}(A-56)\frac{[Y_A-Y_A(0)]}{Y_{56}(0)},$$ where $Y_A$ is the mole fraction of the species with mass number A at the end of the s-process, $Y_A(0)$ the initial mole fraction of the same nucleus, and $Y_{56}(0)$ the initial \chem{56}{Fe} mole fraction (the abundances are averaged over the convective He-burning core);
 \item[-] the maximum convection zone mass extension (hereafter m.c.z.m.e.) during the core He-burning s-process;
 \item[-] the duration of core He-burning s-process (value in the sixth column of Tables \ref{tab:alpha1.7} and \ref{tab:alpha2.0}, and value in the fifth column of Table \ref{tab:old_model}).
\end{itemize}
The results of our computations are summarised in terms of the previous parameters in Tables \ref{tab:alpha1.7} and \ref{tab:alpha2.0} for the mixing length parameters $\alpha=1.7$ and $2.0$,  respectively. The results of s-process test simulation performed with data used in Papers I and II are reported in Table \ref{tab:old_model}. Figures \ref{fig:Fi-a1.7} and \ref{fig:Fi-a2.0} show the overproduction factors of the $6$ s-only nuclei within the mass range $70\leq A\leq 90$ for $\alpha=1.7$ and $2.0$.\par
\begin{table*}
  \caption{Values of the ``indicators'' of the s-process efficiency (see text) for stellar models with mixing length parameter $\alpha=1.7$. The overshooting parameter value used in the stellar evolution code is reported in the first column}
  \label{tab:alpha1.7}
  \centering
  \begin{tabular}{l c c c c c c}
  \hline\hline
  $f$       &$F_0$    &$\tilde{F_0}$ &$A_{max}$  &$n_c$  &$m.c.z.m.e.$   &$Duration$ [$sec$] \\
  \hline
  $10^{-5}$ &$99.88 $ &$6.57 $       &$91-96 $   &$4.24$ &$5.31M_{\odot}$&$2.43\cdot10^{13}$ \\
  $0.01$    &$246.13$ &$11.16$       &$94-104$   &$5.22$ &$6.39M_{\odot}$&$2.27\cdot10^{13}$ \\
  $0.015$   &$216.74$ &$10.13$       &$94-104$   &$5.09$ &$6.87M_{\odot}$&$2.09\cdot10^{13}$ \\
  $0.02$    &$257.81$ &$11.57$       &$94-104$   &$5.32$ &$7.23M_{\odot}$&$2.06\cdot10^{13}$ \\
  $0.025$   &$236.40$ &$10.82$       &$94-104$   &$5.20$ &$7.31M_{\odot}$&$2.16\cdot10^{13}$ \\
  $0.03$    &$304.17$ &$13.20$       &$96-104$   &$5.55$ &$7.97M_{\odot}$&$2.48\cdot10^{13}$ \\
  $0.035$   &$310.39$ &$13.49$       &$96-104$   &$5.58$ &$8.67M_{\odot}$&$2.44\cdot10^{13}$ \\
 \hline
 \end{tabular}
\end{table*}
\begin{table*}
 \caption{Same as Table \ref{tab:alpha1.7} but for stellar models with mixing length parameter $\alpha=2.0$.}
 \label{tab:alpha2.0}
 \centering
 \begin{tabular}{l c c c c c c c}
 \hline\hline
 $f$        &$F_0$    &$\tilde{F_0}$ &$A_{max}$  &$n_c$  &$m.c.z.m.e.$   &$Duration$ [$sec$] \\
 \hline
 $10^{-5}$  &$136.74$ &$7.41 $       &$91-96 $   &$4.48$ &$5.33M_{\odot}$&$2.40\cdot10^{13}$ \\
 $0.01$     &$331.44$ &$14.29$       &$96-110$   &$5.61$ &$6.59M_{\odot}$&$2.32\cdot10^{13}$ \\
 $0.015$    &$222.21$ &$10.28$       &$94-104$   &$5.13$ &$6.74M_{\odot}$&$2.11\cdot10^{13}$ \\
 $0.02$     &$259.42$ &$11.57$       &$94-110$   &$5.31$ &$7.16M_{\odot}$&$2.07\cdot10^{13}$ \\
 $0.025$    &$264.23$ &$11.81$       &$94-110$   &$5.36$ &$7.38M_{\odot}$&$2.09\cdot10^{13}$ \\
 $0.03$     &$309.26$ &$13.35$       &$96-110$   &$5.59$ &$7.93M_{\odot}$&$2.13\cdot10^{13}$ \\
 $0.035$    &$316.68$ &$13.68$       &$96-110$   &$5.64$ &$8.63M_{\odot}$&$1.97\cdot10^{13}$ \\
 \hline
 \end{tabular}
\end{table*}
\begin{table*}
 \caption{Same as Table \ref{tab:alpha1.7} but for calculations performed with the data used in Papers I and II.}
 \label{tab:old_model}
 \centering
 \begin{tabular}{c c c c c c c}
 \hline\hline
 $F_0$   &$\tilde{F_0}$  &$A_{max}$  &$n_c$  &$m.c.z.m.e.$   &$Duration$ [$sec$] \\
 \hline
 $76.89$ &$4.06$         &$89-91$    &$3.84$ &$5.66M_{\odot}$&$2.42\cdot10^{13}$ \\
 \hline
 \end{tabular}
\end{table*}
\section{Results}
\label{Results}
The data reported in Tables \ref{tab:alpha1.7} and \ref{tab:alpha2.0} show the following main features:
\begin{enumerate}
\item a sudden enhancement of the s-process efficiency when passing from a ``no-overshooting'' model ($f=10^{-5}$) to any model that includes overshooting, with the following enhancements for the s-process indicators: $F_0$ and $\tilde{F_0}$ enhanced by a factor $\sim 2\mbox{-}3$ and $\sim 1.5\mbox{-}2$,  respectively; $n_c$ increased from values of $4.2\mbox{-}4.4$ to values of  $5.2\mbox{-}5.6$; $A_{max}$ enhanced from a value around $90$ to values of $104\mbox{-}110$, at least if a minimum overproduction factor of $5$ is used to define $A_{max}$. Even the minimum value adopted for the overshooting parameter ($f=0.01$) is enough to produce a significant increase in all the adopted s-process indicators.
\item a slow change of s-process efficiency when passing from one $f$ value to the next. This is shown more clearly in Figs. \ref{fig:Fi-a1.7} and \ref{fig:Fi-a2.0}, where the overproduction factors for the lightest six s-nuclei are shown for the different $f$ values. For instance, the biggest change in $F_0$ is about a $50\%$ enhancement from the lowest to the highest value for each of the two data sets ($\alpha=1.7$ and $\alpha=2.0$).
\item some anomalies concerning the pattern of the s-process efficiency, which does not seem to grow monotonically strictly with the chosen $f$ values. For each of the two data sets, the main anomaly involves the model with $f=0.01$, which gives birth to a more efficient s-process than the one obtained with higher $f$ values. A minor anomaly is noticed for the model with $f=0.02$ and $\alpha=1.7$ as well.
\end{enumerate}
\section{Discussion}
\label{Discussion}

\subsection{Impact of overshooting}
\label{impact_of_overshooting}
The above results clearly show the level of uncertainty in the modelling of the weak s-process component due to the current lack of a self-consistent theory of stellar convection. More specifically,  the current uncertainties due to overshooting actually do give rise to a factor $\sim 2\mbox{-}3$ uncertainty in the s-process efficiency, an uncertainty that is linked more to the presence or absence ($f=10^{-5}$) of overshooting than to the exact value of the overshooting parameter. This situation would perhaps change in post-He-burning phases.\par
The not strictly monotonic increase in the s-process efficiency with the $f$ value (point 3. Sect. \ref{Results}), may be due to the non-linear nature of the stellar structure and evolution equations. Looking at the values reported in Tables \ref{tab:alpha1.7} and \ref{tab:alpha2.0}, it appears that the m.c.z.m.e. (and the convection zone mass extension, in general) grows monotonically with $f$, but the same is not true for the He-burning phase lifetime.
It can be noticed that the pattern followed by the He-burning lifetime values roughly follows the one of the $F_0$ and $n_c$ values for both data sets. This could be interpreted as follows: the extension of the convection zone has a direct impact on the amount of material available for He-burning and consequently on the s-process duration and neutron exposure. Thus, the longer duration of the s-process for the models with $f=0.01$ (compared to other models with overshooting) is probably the reason for the enhancement of  s-process efficiency.\par
\begin{figure}
  \centering
  \includegraphics[height=8cm]{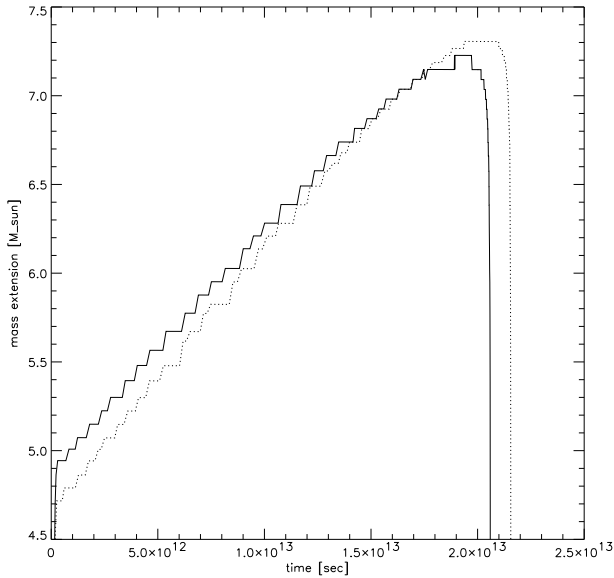}
  \caption{Mass extension of the convection zone (in unit of $M_{\odot}$) as a function of time during the core He-burning s-process for the $\alpha=1.7,~f=0.02$ model (solid line) and for the $\alpha=1.7,~f=0.025$ model (dotted line).}
  \label{fig:convection_zone extension}
\end{figure}

Other anomalies are present in our data. The efficiency of the $\alpha=1.7,~f=0.025$ model is lower compared to the one of the $\alpha=1.7,~f=0.02$ model, despite the fact that corresponding lifetimes follow a ``correctly'' increasing order.
This may be due to a peculiarity of this stellar model (see Fig. \ref{fig:convection_zone extension}): the $f=0.025$ model has a higher m.c.z.m.e., but for most of the time the  $f=0.02$ model shows a larger convection zone. The s-process efficiency reflects this feature, too.
\subsection{Possible contribution from subsequent burning stages}
\label{subsequent_burning_stages}
Arcoragi et al. (\cite{arcoragi91}) suggest that core C-burning and shell He-burning should not cause significant changes to the heavy ($A\gtrsim 60$) nuclide abundances, but the models they developed do not include overshooting.
Other authors (Raiteri et al. \cite{raiteri93}, The et al. \cite{theetal1}, Hoffman et al. \cite{hoffman01}, Woosley et al. \cite{woosley02}) suggest that shell C-burning and other subsequent phases might significantly modify the heavy nuclide abundances.
According to Tables \ref{tab:alpha1.7} and \ref{tab:alpha2.0} (see the m.c.z.m.e. value), the adopted $f$ value does have a big impact on the maximum extension of the convection zone, and a value near $9\, M_{\odot}$ is obtained for $f=0.035$. This might bring s-nuclei that are already produced and some \chem{22}{Ne} towards layers that could be engulfed into subsequent shell-burning stages. Some additional investigation of the possible contribution from post-core-He-burning stages should be performed, taking  uncertainties from stellar evolution and nuclear physics into account. We are convinced that uncertainties in stellar evolution calculations, particularly those concerning post-He burning phases, do not allow any final conclusion about the possible contribution of post-He burning phases to the s-process yields.

\begin{figure}
  \centering
  \includegraphics[height=8cm]{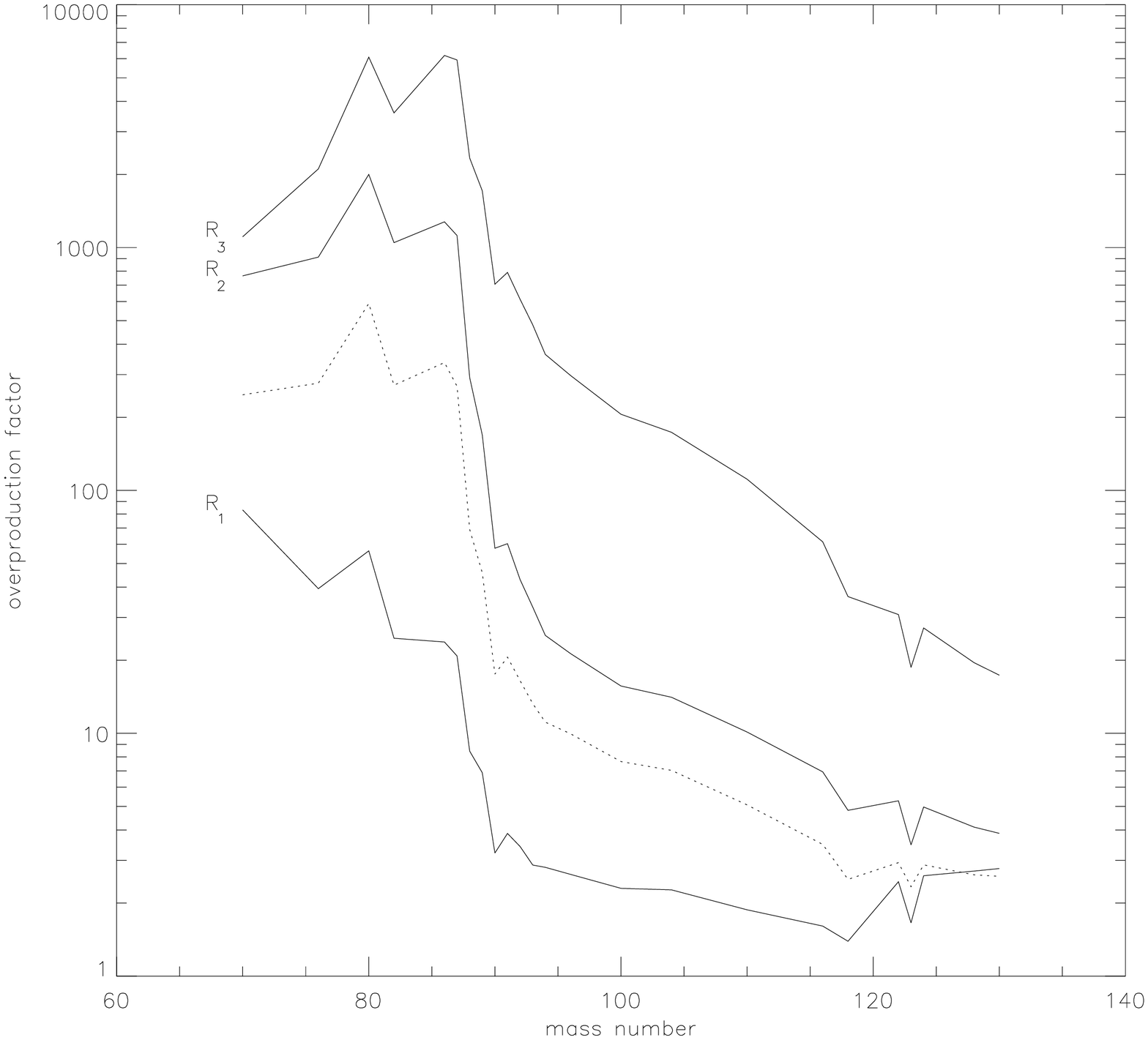}
  \caption{Overproduction factors for the s-nuclei within the mass range $70\leq A\leq 130$. Solid lines show results coming from Paper I, while the dotted line shows our results for the $\alpha=2.0,~f=0.01$ model. The labels $R_i$ (i=1 to 3) refer to the value of the \reac{22}{Ne}{\alpha}{n}{25}{Mg} reaction rate used for s-process calculations in Paper I (see Paper I for the definition of the various $R_i$ values).}
  \label{fig:confronto-costa00}
\end{figure}

\subsection{Impact on the p process}
\label{p_process}
A comparison of our s-process results with those of Paper I (see Fig. \ref{fig:confronto-costa00}) shows that our most efficient s-process is situated between the results obtained in Paper I with the rate $R_1$ and the ones with the rate $R_2$, which means that our results only seem to help a little  for the well-known problem of the underproduction of light p-isotopes in current p-process models (Paper I, Rayet et al. \cite{rayet3}). However, given the fact that the argument for the possible enhancement of the \reac{22}{Ne}{\alpha}{n}{25}{Mg} reaction rate is still being discussed (Jaeger et al. \cite{jaeger01}, Koehler \cite{koehler02}), a combination of a ``moderately'' enhanced s-process (referring to what would be needed for the p-process model to be satisfactory) by means of a stellar model based on overshooting, and a slightly enhanced \reac{22}{Ne}{\alpha}{n}{25}{Mg} reaction rate (like the $R_2$ value reported in Paper I) might possibly bring the light p-isotopes abundances coming from the p-process type II supernova model in line with the solar distribution.

\section{Summary and final comments}
\label{Summary and final comments}
Many s-process simulations have been performed so far by various authors, but there is still a lack of a detailed scrutiny of the impact of stellar evolution modelling uncertainties on the s-process weak component. In this paper we have shown the role of overshooting on the s-process in massive stars.\par
Stellar models of $25\, M_{\odot}$ with initial $Z=0.02$ metallicity have been developed until He exhaustion in the core, using a range of values for the overshooting parameter $f$. Then models were used to ``post-process'' the stellar nuclidic composition with a new s-nucleosynthesis code.\par
Models with overshooting give higher s-process efficiency compared to no-overshooting models, as concluded by Langer et al. (\cite{langer89}) with their simplified treatment of overshooting. However, our enhancements in the average overproduction (factors of about $2$ to $3$) are slightly higher than theirs (factors of about $1.5$ to $2$), and similar behaviour is found for the corresponding $n_c$ values ($4.2\mbox{-}5.6$ and $3.6\mbox{-}4.5$, respectively), although their stellar model treats a $30\, M_{\odot}$ star, which should be expected to give a more efficient s-process than our $25\, M_{\odot}$ stellar models (Prantzos et al. \cite{prantzos90}; The et al. \cite{theetal1}).\par
Less important, but not negligible, variations are obtained when changing the overshooting parameter $f$ value in the $0.01-0.035$ range, and the link between the $f$ value and the s-process indicators values is nearly monotonic.\par
As regards the connection between s- and p-process, the increase in efficiency of the s-process might possibly help with the problem of underproduction with respect to solar of light p-isotopes in type II supernovae.\par
Other studies of stars with different mass and metallicity are needed to confirm our conclusions.\par

\begin{acknowledgements}
The authors wish to thank H. Schlattl for useful suggestions for using of the stellar evolution code Star2003, and M. Arnould for valuable suggestions and a critical reading of an early draft of this manuscript.
\end{acknowledgements}


\end{document}